\documentclass[10pt,a4paper]{article}

\usepackage[utf8]{inputenc}
\usepackage[english]{babel}

\begin{document}
\large

\newpage
\begin{center}
\LARGE{\bf Nature itself in a mirror space-time}
\end{center}
\vspace{0.1mm}
\begin{center}
{\bf Rasulkhozha S. Sharafiddinov}
\end{center}
\vspace{0.1mm}
\begin{center}
{\bf Institute of Nuclear Physics, Uzbekistan Academy of Sciences,
\\Tashkent, 100214 Ulugbek, Uzbekistan}
\end{center}
\vspace{0.1mm}

\begin{center}
{\bf Abstract}
\end{center}

The unity of the structure of matter fields with flavor symmetry laws involves that 
the left-handed neutrino in the field of emission can be converted into a right-handed 
one and vice versa. These transitions together with classical solutions of the Dirac equation 
testify in favor of the unidenticality of masses, energies, and momenta of neutrinos of the 
different components. If we recognize such a difference in masses, energies, and momenta, 
accepting its ideas about that the left-handed neutrino and the right-handed antineutrino 
refer to long-lived leptons, and the right-handed neutrino and the left-handed antineutrino 
are short-lived fermions, we would follow the mathematical logic of the Dirac equation in 
the presence of the flavor symmetrical mass, energy, and momentum matrices. From their point 
of view, nature itself separates Minkowski space into left and right spaces concerning 
a certain middle dynamical line. Thereby, it characterizes any Dirac particle both by left and 
by right space-time coordinates. It is not excluded therefore that whatever the main purposes each 
of earlier experiments about sterile neutrinos, namely, about right-handed short-lived neutrinos 
may serve as the source of facts confirming the existence of a mirror Minkowski space-time. 

\vspace{0.8cm}
\noindent
{\bf 1. Introduction}
\vspace{0.4cm}

A notion about neutrinos introduced by Pauli may be based logically on the availability in nature
of an unbroken flavor symmetry [1]. From its point of view, each type $(l=e,$ $\mu,$ $\tau, ...)$ 
of charged lepton has his own $(\nu_{l}=\nu_{e},$ $\nu_{\mu},$ $\nu_{\tau}, ...)$ neutrino. Such 
pairs are united in families of a definite [2,3] flavor
\begin{equation}
L_{l}=\left\{ {\begin{array}{l}
{+1\quad \mbox{for}\quad l_{L} \, \, \, \, \, l_{R} \, \, \, \,
\nu_{lL} \, \, \, \, \nu_{lR}}\\
{-1\quad \mbox{for}\quad \overline{l}_{R} \, \, \, \, \overline{l}_{L}
\, \, \, \, {\bar \nu_{lR}} \, \, \, \, {\bar \nu_{lL}}}\\
{\, \, \, \, \, 0\quad \mbox{for}\quad \mbox{remaining particles}}\\
\end{array}}\right.
\label{1}
\end{equation}
confirming that nature itself testifies [4,5] in favor of a flavor symmetrical connection 
between the structural particles in difermions 
\begin{equation}
(l_{L}, \overline{l}_{R}) \, \, \, \,
(l_{R}, \overline{l}_{L})
\label{2}
\end{equation}
\begin{equation}
(\nu_{lL}, {\bar \nu_{lR}}) \, \, \, \,
(\nu_{lR}, {\bar \nu_{lL}})
\label{3}
\end{equation}

This in turn implies that the left (right)-handed neutrino in the field of emission similarly to 
a kind of charged lepton [6] can be converted into a right (left)-handed one without change of 
flavor [7]. Such transitions, however, encounter many problems connected with properties of 
a chiral invariance and a mirror symmetry of the best known types of Dirac fermions. 

They reflect the availability of so far unobserved characteristic feature of a latent structure of mass, energy, and momentum and thereby require in principle to fundamentally change our presentations about matter fields. Without such a change, the unified field theory construction of elementary particles still remains not quite in line with nature.

Therefore, we consider the question as to whether there exists any mass dependence 
of spin nature and, if so, what the expected connection says about the dynamical origination of spontaneous mirror symmetry violation. Insofar as a chiral invariance problem is concerned, the results following from its consideration call for special presentation.

\vspace{0.8cm}
\noindent
{\bf 2. Helicity criterion for the Dirac equation}
\vspace{0.4cm}

To express the idea more clearly, it is desirable to use the Dirac equation, which for the 
four-component wave function $\psi(t, {\bf x})$ may be written as
\begin{equation}
i\partial_{t}\psi=\hat H\psi
\label{4}  
\end{equation} 
where it has been accepted that
\begin{equation}
\hat H=\alpha \cdot\hat {\bf p}+\beta m
\label{5}
\end{equation}
Here $\hbar=c=1,$ $E=i\partial_{t},$ and ${\bf p}=-i\partial_{{\bf x}},$ the matrices $\alpha,$ $\beta,$ and $\gamma_{5}$ in the form as were suggested by Dirac [8] have the following structure:
\begin{equation}
\alpha= {{0 \ \, \, \, \, \sigma}\choose{\sigma \, \, \, \, \ 0}} \, \, \, \,
\beta={{I \ \, \, \, \, \ 0}\choose{0 \ -I}} \, \, \, \,
\gamma_{5}={{0 \, \, \, \, \ I}\choose{I \ \, \, \, \, \ 0}}
\label{6}
\end{equation}
Among them $I$ is a unity $2\times 2$ matrix, and $\sigma$ are the Pauli spin matrices. 

Such a choice is based historically on the fact [8] that $\alpha$ and $\beta$ give the possibility 
to directly pass from (\ref{5}) to the relationship involving the mass, energy, and momentum
\begin{equation}
E^{2}={\bf p}^{2}+m^{2}
\label{7}
\end{equation}

Furthermore, if neutrinos are of free particles with an energy $E>0$ then
\begin{equation}
\psi=u({\bf p}, \sigma)e^{-ip \cdot{\it x}}
\label{8}
\end{equation}

With these conditions, $\alpha$ and $\beta$ separate the four-component spinor $u$ into two 
two-component spinors. We must, therefore, replace it with
\begin{equation}
u=u^{(r)}=\left[\chi^{(r)}\atop u_{a}^{(r)}\right]
\label{9}
\end{equation}
where an index $a$ distinguishes $u^{(r)}$ and $u_{a}^{(r)}$ from one another.

The two-component spinors $\chi^{(r)}$ and $u_{a}^{(r)}$ reflect just a regularity that 
(\ref{4}) constitutes the two most diverse equations
\begin{equation}
E\chi^{(r)}=(\sigma{\bf p})u_{a}^{(r)}+m\chi^{(r)}
\label{10}
\end{equation}
\begin{equation}
Eu_{a}^{(r)}=(\sigma{\bf p})\chi^{(r)}-mu_{a}^{(r)}
\label{11}
\end{equation}

This united system in turn establishes two more highly important connections
\begin{equation}
u_{a}^{(r)}=\frac{(\sigma{\bf p})}{E+m}\chi^{(r)} \, \, \, \,
\chi^{(r)}=\frac{(\sigma{\bf p})}{E-m}u_{a}^{(r)}
\label{12}
\end{equation}
and thereby describes a situation in which the availability of each of the two classical 
solutions of the Dirac equation equal to
\begin{equation}
u^{(r)}=\sqrt{E+m}\left[\chi^{(r)}\atop \frac{(\sigma{\bf p})}{E+m}\chi^{(r)}\right]
\label{13}
\end{equation}
says in favor of an explicit mass dependence of spin nature. It is fully possible therefore that mirror symmetry may be violated at the expense of a mass of a particle [9] itself.

Many authors state that there is no connection between the mass of the neutrino and its spin nature. The existence of the latter would seem to contradict our observation that the upper (lower) sign of 
a self-value $s=\pm 1$ of the helicity operator $\sigma{\bf p}=s|{\bf p}|$ corresponds to the right (left)-handed neutrino at the definite choice of spin and momentum directions. But, as stated in (\ref{4}), this implication follows from the fact that in the form as it was accepted, the compound structure of the Dirac equation depending on the mass, energy, and momentum is not in a state to 
give a categorical answer to the question of what of the two four-component spinors (\ref{13}) 
together with 
\begin{equation}
\chi^{(1)}=\pmatrix{1\cr 0}, \, \, \, \, \chi^{(2)}=\pmatrix{0\cr 1}
\label{14}
\end{equation}
describes the same left $(s=L=-1)$ or right $(s=R=+1)$ spin state of the fermion. 

It is also relevant to include in the discussion the free antiparticle with 
\begin{equation}
\psi=\nu({\bf p}, \sigma)e^{-ip \cdot{\it x}}
\label{15}
\end{equation}

If we choose its spinor
\begin{equation}
\nu=\nu^{(r)}=\left[\nu_{a}^{(r)}\atop \chi^{(r)}\right]
\label{16}
\end{equation}
at which an index $a$ is responsible for separation of $\nu^{(r)}$ and $\nu_{a}^{(r)}$ 
from one another, for the case $E<0$ when (\ref{4}) is reduced to 
\begin{equation}
|E|\nu_{a}^{(r)}=-(\sigma{\bf p})\chi^{(r)}-m\nu_{a}^{(r)}
\label{17}
\end{equation}
\begin{equation}
|E|\chi^{(r)}=-(\sigma{\bf p})\nu_{a}^{(r)}+m\chi^{(r)}
\label{18}
\end{equation}
one can find that
\begin{equation}
\nu^{(r)}=\sqrt{|E|+m}\left[\frac{-(\sigma{\bf p})}{|E|+m}\chi^{(r)}\atop \chi^{(r)}\right]
\label{19}
\end{equation}

Comparison of (\ref{19}) with (\ref{13}) at $r=1,2$ leads us to the choice once more of the 
sign of a self-value of quantum mechanical operator $\sigma{\bf p},$ namely, of the sign of 
a helicity $s,$ confirming that we cannot establish the spin nature of elementary particles 
until an equation itself of their unified field theory is able to separate these particles
by the mirror symmetry laws.

\vspace{0.8cm}
\noindent
{\bf 3. Mass, energy, and momentum matrices}
\vspace{0.4cm}

The circumstance in the preceding section seems to indicate that each of transitions
\begin{equation}
\nu_{lL}\leftrightarrow \nu_{lR} \, \, \, \, {\bar \nu}_{lR}\leftrightarrow {\bar \nu}_{lL}
\label{20}
\end{equation}
may serve as the group of arguments in favor of the unidenticality of masses, energies, and momenta 
of neutrinos of the different components without change of their lepton flavors. If we recognize this difference in masses, energies, and momenta, accepting its ideas about that the left-handed neutrino and the right-handed antineutrino refer to long-lived leptons, and the right-handed neutrino and the left-handed antineutrino are of short-lived fermions, we would follow the mathematical logic of the Dirac equation from the point of view of the flavor symmetrical mass, energy, and momentum matrices
\begin{equation}
m_{s}={{m_{V} \, \, \, \, 0}\choose{\ 0 \, \, \, \, \ m_{V}}} \, \, \, \,
E_{s}={{E_{V} \, \, \, \, 0}\choose{\ 0 \, \, \, \, \ E_{V}}} \, \, \, \,
{\bf p}_{s}={{{\bf p}_{V} \, \, \, \, 0}\choose{\ 0 \, \, \, \, \ {\bf p}_{V}}}
\label{21}
\end{equation}
\begin{equation}
m_{V}={{m_{L} \, \, \, \, 0}\choose{\ 0 \, \, \, \, \ m_{R}}} \, \, \, \,
E_{V}={{E_{L} \, \, \, \, 0}\choose{\ 0 \, \, \, \, \ E_{R}}} \, \, \, \,
{\bf p}_{V}={{{\bf p}_{L} \, \, \, \, 0}\choose{\ 0 \, \, \, \, \ {\bf p}_{R}}}
\label{22}
\end{equation}
where $V$ must be considered as an index of a distinction.

At these situations, any of interconversions (\ref{20}) can be explained by the particles 
$\nu_{lL}({\bar \nu_{lR}})$ and $\nu_{lR}({\bar \nu_{lL}})$ possessing unidentical masses, 
energies, and momenta. But their Coulomb nature has been created so that to each type of 
C-even or C-odd charge corresponds a kind of current [5]. There exists, therefore, the possibility 
that a classification of leptonic currents with respect to C-operation admits the existence of two types of leptons of vector $V_{l}$ and axial vector $A_{l}$ currents of different C-invariance [10]. Unlike the fermions of a C-odd charge, the mass of which is strictly an axial vector $(A)$ type, 
the particles of a C-even charge have mass of a vector $(V)$ nature [5]. 

It is already clear from the foregoing that (\ref{21}) and (\ref{22}) refer to those neutrinos among which there are no elementary objects with axial vector masses, energies and momenta.

In the presence of such matrices, the structure of the unified field theory equation of neutrinos 
of vector types becomes fully definite and behaves as follows:
\begin{equation}
i\frac{\partial}{\partial t_{s}}\psi_{s}=\hat H_{s}\psi_{s}
\label{23}  
\end{equation} 
in which 
\begin{equation}
\hat H_{s}=\alpha \cdot\hat {\bf p_{s}}+\beta m_{s}
\label{24}
\end{equation}
and $E_{s}$ and ${\bf p}_{s}$ correspond to quantum energy and momentum operators 
\begin{equation}
E_{s}=i\frac{\partial}{\partial t_{s}} \, \, \, \,  
{\bf p}_{s}=-i\frac{\partial}{\partial {\bf x}_{s}}
\label{25}
\end{equation}

As well as in (\ref{21}), the index $s$ here expresses, as we see later, the unidenticality 
of space-time coordinates $[(t_{s}, {\bf x}_{s})]$ of the left- and right-handed particles. Then 
it is possible, for example, to describe the field of the free neutrino in a latent united form
\begin{equation}
\psi_{s}=u_{s}({\bf p}_{s}, \sigma)e^{-ip_{s} \cdot{\it x}_{s}} \, \, \, \, E_{s}>0
\label{26}
\end{equation}

Because of (\ref{6}), (\ref{21}), and (\ref{22}), the four-component wave function 
$\psi_{s}(t_{s}, {\bf x}_{s})$ is reduced at first to the two two-component wave functions 
and, next, the latter separates it into four possible parts. Formulating more concretely, 
one can write the field $u_{s}$ in a general form
\begin{equation}
u_{s}=u^{(r)}=\left[\chi^{(r)}\atop u_{a}^{(r)}\right]
\label{27}
\end{equation}

So, we must recognize that (\ref{23}) together with (\ref{6}), (\ref{21}), (\ref{26}), and 
(\ref{27}) constitutes the naturally united system of Dirac equations
\begin{equation}
E_{V}\chi^{(r)}=(\sigma{\bf p}_{V})u_{a}^{(r)}+m_{V}\chi^{(r)}
\label{28}
\end{equation}
\begin{equation}
E_{V}u_{a}^{(r)}=(\sigma{\bf p}_{V})\chi^{(r)}-m_{V}u_{a}^{(r)}
\label{29}
\end{equation}

Their two-component spinors correspond to the fact that in them, $m_{V},$ $E_{V},$ and ${\bf p}_{V}$ are the flavor symmetrical $2\times 2$ matrices, which are absent in a classical system of Dirac equations. Instead they include the usual mass, energy, and momentum.

It is not surprising therefore that at the availability of a connection
\begin{equation}
u_{a}^{(r)}=\frac{(\sigma{\bf p}_{V})}{E_{V}+m_{V}}\chi^{(r)} \, \, \, \,
\chi^{(r)}=\frac{(\sigma{\bf p}_{V})}{E_{V}-m_{V}}u_{a}^{(r)}
\label{30}
\end{equation}
any of the two solutions of the new Dirac equation (\ref{23}) equal to
\begin{equation}
u^{(r)}=\sqrt{E_{V}+m_{V}}
\left[\chi^{(r)}\atop \frac{(\sigma{\bf p}_{V})}{E_{V}+m_{V}}\chi^{(r)}\right]
\label{31}
\end{equation}
respond to the same left- or right-handed neutrino.

To investigate further, we present (\ref{31}) in an explicit form 
\begin{equation}
u^{(1)}=\sqrt{E_{L}+m_{L}}
\left[
\begin{array}{c}
1\\ 0\\ \frac{(\sigma{\bf p}_{L})}{E_{L}+m_{L}}\\ 0
\end{array}
\right]
\label{32}
\end{equation}
\begin{equation}
u^{(2)}=\sqrt{E_{R}+m_{R}}
\left[
\begin{array}{c}
0\\ 1\\ 0\\ \frac{(\sigma{\bf p}_{R})}{E_{R}+m_{R}}
\end{array}
\right]
\label{33}
\end{equation}

It is seen that $u^{(1)},$ $\chi^{(1)},$ and $u_{a}^{(1)}$ characterize the left-handed 
neutrino, and $u^{(2)},$ $\chi^{(2)},$ and $u_{a}^{(2)}$ describe the right-handed neutrino.

For completeness we include in the discussion the free antineutrino with 
\begin{equation}
\psi_{s}=\nu_{s}({\bf p}_{s}, \sigma)e^{-ip_{s} \cdot{\it x}_{s}} \, \, \, \, E_{s}<0
\label{34}
\end{equation}

At the same time, (\ref{6}) and (\ref{21}) replace the spinor $\nu_{s}$ for
\begin{equation}
\nu_{s}=\nu^{(r)}=\left[\nu_{a}^{(r)}\atop \chi^{(r)}\right]
\label{35}
\end{equation}
and thereby transform (\ref{23}) into the two other equations 
\begin{equation}
|E_{V}|\nu_{a}^{(r)}=-(\sigma{\bf p}_{V})\chi^{(r)}-m_{V}\nu_{a}^{(r)}
\label{36}
\end{equation}
\begin{equation}
|E_{V}|\chi^{(r)}=-(\sigma{\bf p}_{V})\nu_{a}^{(r)}+m_{V}\chi^{(r)}
\label{37}
\end{equation}

By following the same arguments that led to (\ref{19}), but having
in view the equality
\begin{equation}
\nu^{(r)}=\sqrt{|E_{V}|+m_{V}}
\left[\frac{-(\sigma{\bf p}_{V})}{|E_{V}|+m_{V}}\chi^{(r)}\atop \chi^{(r)}\right]
\label{38}
\end{equation}
one can also make a conclusion that 
\begin{equation}
\nu^{(1)}=\sqrt{|E_{L}|+m_{L}}
\left[
\begin{array}{c}
\frac{-(\sigma{\bf p}_{L})}{|E_{L}|+m_{L}}\\ 0\\ 1\\ 0
\end{array}
\right]
\label{39}
\end{equation}
\begin{equation}
\nu^{(2)}=\sqrt{|E_{R}|+m_{R}}
\left[
\begin{array}{c}
0\\ \frac{-(\sigma{\bf p}_{R})}{|E_{R}|+m_{R}}\\ 0\\ 1
\end{array}
\right]
\label{40}
\end{equation}
which show that $\nu^{(1)},$ $\chi^{(1)},$ and $\nu_{a}^{(1)}$ correspond to the right-handed antineutrino, and $\nu^{(2)},$ $\chi^{(2)},$ and $\nu_{a}^{(2)}$ respond to the left-handed antineutrino.

Thus, we have established the full spin structure of the Dirac equation in which it is 
definitely stated that
\begin{equation}
\sigma{\bf p_{L}}=-|{\bf p}_{L}| \, \, \, \, \sigma{\bf p_{R}}=|{\bf p}_{R}|
\label{41}
\end{equation}

Simultaneously, as is easy to see, the neutrino $\nu_{lL}$ and the antineutrino ${\bar \nu}_{lR}$
are the left-polarized leptons, and the neutrino $\nu_{lR}$ and the antineutrino ${\bar \nu}_{lL}$
refer to the right-polarized fermions. 

In these circumstances, it seems possible to use $\psi_{s}$ in the form
\begin{equation}
\psi_{s}=\pmatrix{\psi\cr \phi} \, \, \, \,
\psi=\pmatrix{\psi_{L}\cr \psi_{R}} \, \, \, \,
\phi=\pmatrix{\phi_{L}\cr \phi_{R}}
\label{42}
\end{equation}

Uniting (\ref{42}) with (\ref{23}) and solving the finding equations concerning $\psi_{L,R}$
and $\phi_{L,R},$ it can also be verified that
\begin{equation}
E_{L,R}^{2}={\bf p}_{L,R}^{2}+m_{L,R}^{2}
\label{43}
\end{equation}

The difference in lifetimes of neutrinos of the different components can explain the spontaneous 
mirror symmetry violation, at which they have unidentical masses, energies, and momenta. 
This leads us to the conclusion that
\begin{equation}
E_{L}=i\frac{\partial}{\partial t_{L}} \, \, \, \,  
E_{R}=i\frac{\partial}{\partial t_{R}} 
\label{44}
\end{equation}
\begin{equation}
{\bf p}_{L}=-i\frac{\partial}{\partial {\bf x}_{L}} \, \, \, \,  
{\bf p}_{R}=-i\frac{\partial}{\partial {\bf x}_{R}}
\label{45}
\end{equation}

Insofar as the mass is concerned, we start from the special comparison theorem [11] for the Dirac equation that $m_{s}$ similarly to all $E_{s}$ and ${\bf p}_{s}$ must be quantum operators such as 
\begin{equation}
m_{L}=-i\frac{\partial}{\partial \tau_{L}} \, \, \, \,  
m_{R}=-i\frac{\partial}{\partial \tau_{R}} 
\label{46}
\end{equation}
where $\tau_{L}$ and $\tau_{R}$ are the lifetimes of the left- and right-handed 
particles, respectively. 

Furthermore, if these situations are of fundamental principles of quantum mechanics, our reasoning refers to all the particles interacting according to (\ref{23}), which unites the massive Dirac fermions of vector types.

Such a connection arises as a consequence of the ideas of corresponding mechanism laws responsible for the dynamical origination of spontaneous mirror symmetry violation. From their point of view, 
nature itself separates Minkowski space into left and right spaces concerning a certain middle dynamical line. Thereby, it characterizes any Dirac particle both by left $[(t_{L}, {\bf x}_{L})]$ and by right $[(t_{R}, {\bf x}_{R})]$ space-time coordinates. In this it is additionally assumed 
that $\tau_{L}$ and $\tau_{R}$ correspond in (\ref{46}) to the lifetimes of a particle in the 
left and right Minkowski spaces, respectively. It is not excluded therefore that whatever the 
main purposes the recent experiments [12] about sterile neutrinos, namely, about right-handed 
short-lived neutrinos may serve as the first confirmation of the existence of 
a mirror Minkowski space-time. 

\vspace{0.8cm}
\noindent
{\bf 4. Conclusion}
\vspace{0.4cm}

There exists of course a range of old phenomena in which appears a part of the dynamical 
origination of spontaneous mirror symmetry violation. A beautiful example is the solar 
neutrino problem [13].

At first sight, an active left-handed neutrino passing through the medium from the Sun to a detector 
on the Earth can be converted into a sterile right-handed neutrino [14] not interacting with the field of emission, for example, in the reactions $\nu_{eL,R}+Cl^{37}\rightarrow e_{L,R}+Ar^{37}$ 
as well as in other phenomena with neutrinos. Therefore, it seems that an observed flux of neutrinos will be half that of the starting one. However, as we shall see, this is not quite so. The point is that the left-handed long-lived neutrino at the interaction with matter will be converted into a right-handed short-lived neutrino without change of its lepton flavor. The right-handed neutrino 
in turn interacts with the field of emission until it will not virtually decay forming the real 
left-handed neutrino of the same flavor. Under such circumstances, a flux of solar neutrinos 
does not suffer a decrease in quantity.

In the standard electroweak model [15-17], it has been usually assumed that in nature the 
right-handed neutrino is absent. This is of course intimately connected with the prediction 
of a two-component theory [18] of the neutrino, expressing the idea of parity nonconservation in 
the weak interactions [19]. According to one of its aspects, the matrix $\gamma_{5}$ in the chiral presentation of the Weyl [20] constitutes the projection operator $(1-\gamma_{5})/2$ allowing one 
to choose only the left components of the four-component spinor.

Such a procedure, however, redoubles the results of theoretical calculations in all flavor symmetrical processes with weak charged currents even in the presence of a normalized multiplier. Insofar as the weak neutral currents are concerned, the terms with $\gamma_{5}$ appear in them as 
the axial vector components of these currents. The number of solar neutrinos and the structural phenomena originating in the detectors on the Earth coincide, as follows from considerations of flavor symmetry. This conformity requires comparison with experiment of any of the two equal 
parts of a theoretical estimate of a flux of solar neutrinos.

Thus, if the structure of the standard model is not quite in line with ideas of a new Dirac 
equation, (\ref{23}), it needs in special reconstruction.

Finally, insofar as the Dirac Lagrangian and its structural components are concerned, all 
of them together with some aspects of spontaneous mirror symmetry violation (not noted here)
will be presented in the separate work.
\vspace{0.8cm}

\noindent
{\bf References}
\begin{enumerate}
\item
R.S. Sharafiddinov. Bull. Am. Phys. Soc. {\bf 59.} L1.00036 (2014). 
\item
Ya.B. Zel'dovich. Dokl. Akad. Nauk SSSR. {\bf 91.} 1317 (1953).
\item
E.J. Konopinski and H. Mahmoud. Phys. Rev. {\bf 92.} 1045 (1953).

doi: 10.1103/PhysRev.92.1045. 
\item
R.S. Sharafiddinov. Bull. Am. Phys. Soc. {\bf 59.} T1.00004 (2014).
\item
R.S. Sharafiddinov. Fiz. {\bf B. 16.} 1 (2007). 2005. arXiv:hep-ph/0512346.
\item
R.S. Sharafiddinov. Eur. Phys. J. Plus {\bf 126.} 40 (2011). 

doi:10.1140/epjp/i2011-11040-x.
\item
R.S. Sharafiddinov. Can. J. Phys. {\bf 92.} 1262 (2014). 

doi: 10.1139/cjp-2013-0458.
\item
P.A.M. Dirac. Proc. Roy. Soc. Lond. {\bf A 117.} 610 (1928). doi: 10.1098/rspa.1928.0023. 
\item
R.S. Sharafiddinov. Phys. Essays {\bf 19.} 58 (2006). 
\item
R.S. Sharafiddinov. Bull. Am. Phys. Soc. {\bf 57.} KA.00069 (2012). 2010.

arXiv:1004.0997 [hep-ph].
\item
R. L. Hall. Phys. Rev. Lett. {\bf 101.} 090401 (2008). doi: 10.1103/PhysRevLett.101.090401. 
\item
P. Adamson, et al. Phys. Rev. Lett. {\bf 110.} 251801 (2013). 

doi: 10.1103/PhysRevLett.110.251801.
\item
R. Davis, Jr. Phys. Rev. Lett. {\bf 12.} 303 (1964). doi: 10.1103/PhysRevLett.12.303 
\item
L.B. Okun. Yad. Fiz. {\bf 44.} 847 (1986).
\item
S.L. Glashow. Nucl. Phys. {\bf 22.} 579 (1961). doi: 10.1016/0029-5582(61)90469-2.
\item
A. Salam and J.C. Ward. Phys. Lett. {\bf 13.} 168 (1964). doi: 10.1016/0031-9163(64)90711-5.
\item
S. Weinberg. Phys. Rev. Lett. {\bf 19.} 1264 (1967). doi: 10.1103/PhysRevLett.19.1264.
\item
T.D. Lee and C.N. Yang. Phys. Rev. {\bf 105.} 1671 (1957). doi: 10.1103/PhysRev.105.1671. 
\item
T.D. Lee and C.N. Yang. Phys. Rev. {\bf 104.} 254 (1956). doi: 10.1103/PhysRev.104.254. 
\item
H. Weyl. Z. Phys. {\bf 56.} 330 (1929). doi: 10.1007/BF01339504. 
\end{enumerate}
\end{document}